\documentclass[12pt,preprint]{aastex}

\def\ltsima{$\; \buildrel < \over \sim \;$}
\def\simlt{\lower.5ex\hbox{\ltsima}} 
\def\gtsima{$\; \buildrel > \over \sim \;$}
\def\simgt{\lower.5ex\hbox{\gtsima}} 

\shorttitle{FR~Is and $\gamma$-ray Background} 
\shortauthors{Stawarz, Kneiske \& Kataoka}

\begin{document}

\title{Kiloparsec-Scale Jets in FR~I Radio Galaxies\\ and the $\gamma$-ray Background}

\author{\L . Stawarz\altaffilmark{1}}

\affil{Max-Planck-Institut fur Kernphysik, PO Box 103980, 69029 Heidelberg, Germany \email{Lukasz.Stawarz@mpi-hd.mpg.de}}

\altaffiltext{1}{also at Obserwatorium Astronomiczne, Uniwersytet Jagiello\'nski, ul. Orla 171, 30-244 Krak\'ow, Poland; stawarz@oa.uj.edu.pl}

\author{T. M. Kneiske}

\affil{Department of Physics and Mathematical Physics, University of Adelaide, North Terrace, Adelaide, SA 5005, Australia \email{tkneiske@physics.adelaide.edu.au}}

\author{J. Kataoka}

\affil{Department of Physics, Tokyo Institute of Technology, 2-12-1, Ohokayama, Meguro, Tokyo 152-8551, Japan \email{kataoka@hp.phys.titech.ac.jp}}

\begin{abstract}
  
We discuss the contribution of kiloparsec-scale jets in FR~I radio galaxies to the diffuse $\gamma$-ray background radiation. The analyzed $\gamma$-ray emission comes from inverse-Compton scattering of starlight photon fields by the ultrarelativistic electrons whose synchrotron radiation is detected from such sources at radio, optical and X-ray energies. We find that these objects, under the minimum-power hypothesis (corresponding to a magnetic field of $300$\,$\mu$G in the brightest knots of these jets), can contribute about one percent to the extragalactic $\gamma$-ray background measured by EGRET. We point out that this result already indicates that the magnetic fields in kpc-scale jets of low-power radio galaxies are not likely to be smaller than $10$\,$\mu$G on average, as otherwise the extragalactic $\gamma$-ray background would be overproduced.
 
\end{abstract}

\keywords{radiation mechanisms: non-thermal --- galaxies: active --- galaxies: jets --- gamma-rays: theory}

\section{Introduction}

Observations by the EGRET instrument on board the CGRO have established
presence of isotropic extragalactic background radiation in the $100$\,MeV
-- $10$\,GeV photon energy range, with an integrated flux $I(> 100 \,
{\rm MeV}) \lesssim 10^{-5}$\,ph\,cm$^{-2}$\,s$^{-1}$\,sr$^{-1}$ and a
curved (concave) spectrum \citep{sre98,str04}. It has been argued
\citep{der92,ste93,pad93,sal94,set94,chi95,erl95,ste96,chi98,wef99,muc00},
that the bulk of this emission most likely originates from unresolved
blazars, whose properties are similar to those detected by EGRET
\citep{mon95,har99}. However, as the internal parameters and
cosmological distribution of EGRET-like blazar sources are not
precisely known, the origin of the diffuse $\gamma$-ray background is
still being debated, and new classes of objects emitting GeV photons
can be analyzed in this context \citep[e.g.,][]{loe00,gab03}.

Here we discuss the issue of high-energy $\gamma$-ray emission from
the kiloparsec-scale jets in FR~I radio galaxies, and in particular its
contribution to the $\gamma$-ray background radiation as measured by
EGRET. Our study is stimulated by recent results from the Chandra X-ray
Observatory, which have shown that X-ray jet emission is common in FR~I
sources
\citep{hard01,hard02,hard03,hard05,har02a,har02b,kat03,kra02,mar02,wil02,wor01,wor03}.
The established synchrotron origin of this X-ray emission (see
all the references above\footnote{We note that the origin of the X-ray
emission in powerful large-scale jets in quasars is still under
debate \citep[see the discussion in][and references therein]{kat05},
but the synchrotron scenario for the X-ray FR~I jets is 
widely accepted.}) implies that the kpc-scale jets in FR~I radio
galaxies will, at some level, be sources of high- and very high-energy
$\gamma$-ray emission due to the inverse-Compton scattering of ambient
(galactic) photon fields by the synchrotron-emitting electrons. This
problem was discussed by \citet{sta03}, and considered in more
detail for the specific case of the radio galaxy M~87 by \citet{sta05}.
The aim of our study --- which is an extension of the previous
analysis --- is twofold, namely an estimation of the aforementioned
contribution (taking into account effects of absorption and subsequent
re-processing of $\gamma$-ray photons by infrared-to-ultraviolet
background radiation), and discussion of the constraints on the jet
parameters which can be imposed in this way.

The paper is organized as follows. In the next section we describe
the ingredients of the model developed to estimate the contribution of
kpc-scale FR~I jets to the $\gamma$-ray background. In section~3 we
present the  results, and discuss briefly the possibility of the
direct observation of $\gamma$-rays from these sources.
General discussion and the main conclusions are presented
in section~4. Throughout the paper we assume a cosmology with
$\Omega_{\rm M} = 0.27$, $\Omega_{\Lambda} = 0.73$, and 
$H_0 = 71$\,km\,s$^{-1}$\,Mpc$^{-1}$.

\section{The Model}

Below we describe the procedure used to estimate the $\gamma$-ray emission
from a `typical' kpc-scale jet in a FR~I radio galaxy. We assume that
all of these objects have similar properties, and that the detection
rate of their brightest knots at optical and X-ray frequencies 
depends solely on the amount of relativistic beaming
\citep{spa95,sca02,jes03}. Within this approach, we find the
`universal' electron energy distribution in the brightest knots of
these jets by fitting a broken power-law to the radio--to--X-ray
synchrotron continua of the collected jet sources. Next, we compute the
$\gamma$-ray emission due to inverse-Compton (IC) scattering of the
re-constructed universal electron energy distribution on the starlight
radiation of the host galaxies \citep{sta03,sta05}. 
We include both the Klein-Nishina effects and the relativistic bulk
velocity of the emitting plasma in the analysis. 
We also present the adopted approach
in relating the $\gamma$-ray output of the jets with the
total radio luminosities of the analyzed sources, and hence the
$\gamma$-ray luminosity function (GLF) of kpc-scale FR~I jets with the
radio luminosity function (RLF) of low-power radio galaxies
\citep{wil01}. Finally, we briefly describe the adopted model for
the absorption of the emitted
high-energy $\gamma$-ray photons by the
infrared-to-ultraviolet metagalactic radiation field (MRF), taking into
account evolution of the background photon field up to high redshifts
\citep{kne02,kne04}, and also re-processing of the absorbed
$\gamma$-ray photons to lower energies by the cascading processes.

\subsection{Jet $\gamma$-ray Emission}

We collate data for all FR~I radio galaxies with detected X-ray
jets. We restrict our analysis to the brightest knots in these
jets, which are placed at $\sim 1$ kpc from the galactic nuclei,
thus obtaining a list of 11 sources. Table~1 summarizes the input
parameters for the considered knots, i.e., the radio fluxes $S_{\rm R}$
as measured at $\nu_{\rm R} = 5$\,GHz, the optical fluxes $S_{\rm O}$
(if available) at $\nu_{\rm O} = 5 \times 10^{14}$\,Hz, and the X-ray
fluxes $S_{\rm X}$ at $h \nu_{\rm X} = 1$\,keV photon energies. We fit
broken power-laws to the radio-to-X-ray continua of the analyzed
knots, with spectral indices $\alpha_{\rm R}$ and $\alpha_{\rm X}$
measured directly from the data or, when large uncertainties are
encountered, with the adopted median values $\alpha_{\rm R} = 0.75$
and $\alpha_{\rm X} = 1.2$. We assume a spherical geometry of radius
$26$\,pc for all knots, which corresponds to an angular radius of
$0.3''$ at the distance of M~87 \citep[see in this context][]{kat05}.
The obtained values of the break frequencies in the synchrotron
spectra, $\nu_{\rm br}$, as well as the equipartition magnetic field
computed for no relativistic beaming, $B_{\rm eq}(1)$, are also given in
Table~1, together with the radio luminosity of the whole
source, $L_{\rm tot}$, as given by
\citet{liu02} after conversion to our adopted cosmology and an
observation frequency of 5\,GHz \citep[assuming a radio spectral index 
for the whole source of 0.8; see][]{wil01}. Finally, in Table~1 we
give the estimated value of the ratio between the 5\,GHz luminosity of
each analyzed knot and the 5\,GHz total luminosity of the source,
\begin{equation}
\eta \equiv {4 \pi \, d_{\rm L}^2 \times \left[\nu_{\rm R} S_{\rm R}\right] \over L_{\rm tot}} \quad .
\end{equation}
\noindent

The fitting procedure described above allows us to find the `typical'
parameters of the brightest knots in FR~I radio galaxies, which we
adopt hereafter. Besides the spectral indices, $\alpha_{\rm R} = 0.75$
and $\alpha_{\rm X} = 1.2$, they are the median synchrotron break
frequency, $\nu_{\rm br} = 10^{14}$\,Hz, the median equipartition
magnetic field, $B_{\rm eq}(1) = 300$\,$\mu$G, and the median luminosity
ratio, $\eta \approx 0.02$. Assuming hereafter a universal value of 
$\eta = 0.01$ for all FR~I sources, we are able to express the synchrotron
luminosity of each knot (and hence to normalize the energy distribution of
the synchrotron-emitting electrons for each knot) in terms of the total
radio luminosity of the source. We further choose `typical' values for
the kinematic parameters of the jets at the position of the analyzed
knots, namely a bulk Lorentz factor of $\Gamma_{\rm j} = 3$ and a jet viewing
angle of $\theta = 45^\circ$. This particular (although sufficient for our
purposes) choice is in agreement with the one-sidedness of the X-ray jets
detected in FR~I radio galaxies, and with the jet--counterjet radio
brightness asymmetry for all the sources analyzed in this paper. Also,
it results in a Doppler factor of the emitting plasma of $\delta = 1$,
leading to $B_{\rm eq} = B_{\rm eq}(1)$.

With the above parameters, we reconstruct the energy distribution of the
synchrotron-emitting electrons, and then compute their unabsorbed
$\gamma$-ray emission due to IC scattering on the starlight photon
field. We use formulae from \citet{aha81}, including the Klein-Nishina
effects and relativistic velocity of the emitting plasma \citep[see a
detailed description in][]{sta05}. We assume a characteristic frequency
of the starlight emission of $\nu_{\rm star} = 10^{14}$\,Hz, and a
bolometric starlight energy density at $\sim 1$\,kpc from the galactic
center of $U_{\rm star} = 10^{-9}$ erg cm$^{-3}$, both as measured in the
host-galaxy rest frame \citep[see][]{sta03}. The starlight
radiation energy density then dominates the other photon fields in the
jet rest frame, in particular the energy density of the synchrotron
photons, by more than an order of magnitude for knot synchrotron
luminosities $\leq 10^{42}$\,erg\,s$^{-1}$ and other parameters as
discussed in the paper.

The resulting intrinsic IC luminosity, $L_{\gamma}(\varepsilon)$,
consists of a power-law continuum $\propto \varepsilon^{-0.75}$ at
photon energies $\varepsilon < 1$\,GeV, a break region between $1$\,GeV
and $100$\,GeV due to both the transition between the Thomson and the
Klein-Nishina regimes and a break in the electron energy
distribution, and finally a steep power-law for 
$\varepsilon \geq 100$\,GeV (see figure~1). 
With a maximum electron energy of $E_{\rm e}/m_{\rm e} c^2 = 10^8$ 
assumed hereafter, the maximum IC photon energy is
$\varepsilon \approx 50$\,TeV. Note that the energy density of the jet
magnetic field in the emitting plasma rest frame (denoted by primes),
$U'_{\rm B} = B_{\rm eq}^2 / 8 \pi \approx 3.6 \times 10^{-9}$\,erg\,cm$^{-3}$,
is roughly comparable with the energy density of the
starlight radiation, $U'_{\rm star} = \Gamma_{\rm j}^2 \, U_{\rm star}
\approx 9 \times 10^{-9}$\,erg\,cm$^{-3}$, and hence the intrinsic IC
luminosity is roughly comparable with the synchrotron luminosity. In
particular, by integrating over the IC photon energy range,
$L_{\gamma} = \int_{\varepsilon_{\rm min}}^{\varepsilon_{\rm max}} L_{\gamma}(\varepsilon) \, d \varepsilon$, we find that
\begin{equation}
L_{\gamma} \approx \left({\eta \over 0.01} \right) \times L_{\rm tot}
\end{equation}
\noindent
for $\varepsilon_{\rm min} = 100$\,MeV and $\varepsilon_{\rm max} =
300$\,GeV. Due to the specific spectral shape of the intrinsic IC
emission, with a maximum at $1-100$\,GeV, the above
relation is expected to be correct also for $\varepsilon_{\rm min} <
100$\,MeV and $\varepsilon_{\rm max} > 300$\,GeV. Obviously, the
observed IC emission will be modified by the effects of absorption and
subsequent re-processing of high-energy $\gamma$-ray photons on the
diffuse cosmic background radiation fields, as described below.

\subsection{Absorption and Re-processing of $\gamma$-ray Photons}

High-energy $\gamma$-ray photons produced at cosmological distances
are likely to be absorbed by the diffuse cosmic background radiation
due to photo-photon annihilation \citep{nik62,gou66,ste92}. The
absorbed $\gamma$-rays are then re-emitted at lower energies due to
inverse-Compton radiation of the created electron-positron pairs in
the cascading process \citep{aha85,pro86,zdz88,pro93,aha94,kel04}.
Among other implications, this process is of significant interest
in studying the contribution of cosmologically distant sources to
the $\gamma$-ray background \citep{cop97,kne05}. A crucial point
in the analysis of the attenuation of high-energy $\gamma$-rays
emitted by cosmologically distant objects is, however, the knowledge of
spectral shape of the MRF in the infrared-to-ultraviolet photon energy
range, which is responsible for the $\gamma$-ray
absorption/re-emission, and of its evolution up to high redshifts \citep[see][]{pri99}.
This is a complicated problem, since the direct measurements of 
the MRF at infrared wavelengths are difficult 
\citep[see][and references therein]{hau01}. 
We note that careful analysis of the spectra of TeV-emitting blazars
constitutes a very promising (although not direct, and, in addition,
restricted to redshifts $z<1$) method for constraining unknown
parameters of the MRF \citep[e.g.,][]{sta98,ren01,aha02,cos04,dwe05}.

Here we adopt a model proposed by \citet{kne02,kne04}, which under a minimum
of parameters and assumptions follows the evolution of the MRF 
from redshift $z=5$ to $z=0$, fulfilling all the observational constraints
on the MRF in the infrared-to-optical band at the present epoch. This forward evolution model based on optical and
infrared galaxy surveys considers emission from stars, gas and dust in
optically selected galaxies as well as in luminous and
ultraluminous infrared galaxies (LIGs/ULIGs). The contribution
of LIGs is essential, since the number of stars formed in this object
is comparable to the one in optically selected galaxies, and their
contribution to the infrared-to-ultraviolet metagalactic radiation field
is about 50$\%$.
We note that in a framework of the adopted model the UV background is underestimated at
redshifts $z>3$ by a factor of 2-4. With this for the MRF, we derive the optical depth for
$\gamma$-ray absorption (including the effects of re-emission) as a
function of redshift and $\gamma$-ray photon energy, 
$\tau_{\gamma \gamma}(\varepsilon, \, z)$ \citep[see][]{kne05}, 
and compute the observed absorbed IC flux of the selected FR~I jets,
\begin{equation}
S_{\gamma}(\varepsilon) = {(1+z) \, L_{\gamma}\left[(1+z) \, \varepsilon\right] \over 4 \pi \, d_{\rm L}^2} \times \exp \left[ - \tau_{\gamma \gamma}(\varepsilon, \, z) \right] \quad .
\end{equation}
\noindent
Figure~1 shows a comparison between the unabsorbed and the absorbed IC
fluxes from kpc-scale jets located at different redshifts, $z=0.03$,
$0.1$, $1.0$, and $4.0$ (for $L_{\gamma} = 10^{41}$\,erg\,s$^{-1}$). As
illustrated, only a relatively small part of the IC flux is absorbed by
the MRF in the $0.1-10$\,GeV photon energy range, and hence our
analysis of the contribution of kpc-scale FR~I jets to the
$\gamma$-ray background radiation is not greatly affected by the
inevitably somewhat arbitrary nature of the adopted MRF model.

\subsection{$\gamma$-ray Luminosity Function}

We take the radio luminosity function --- i.e. the number of sources
per unit comoving volume per unit (base $10$) logarithm of the source
luminosity --- characterizing low-power radio sources (including
`classical' FR~I radio galaxies and FR~II radio galaxies with
weak/absent emission lines) in the form:
\begin{equation}
\rho_{\rm W}(L, \, z) \equiv {dN \over dV \, d \log L} = \left\{\begin{array}{ccc} \rho_0 \, \left({L \over L_{\rm cr}}\right)^{-\alpha} \, \exp\left({-L \over L_{\rm cr}}\right) \, (1+z)^k & {\rm for} & z < z_{\rm cr} \\ \rho_0 \, \left({L \over L_{\rm cr}}\right)^{-\alpha} \, \exp\left({-L \over L_{\rm cr}}\right) \, (1+z_{\rm cr})^k & {\rm for} & z \geq z_{\rm cr} \end{array} \right.
\end{equation}
\noindent
as discussed by \citet{wil01}. We adopt the values of the five free parameters
of $\rho_{\rm W}(L, \, z)$ as determined from 3CRR, 6CE and 7CRS
samples for a $\Omega_{\rm M} = \Omega_{\Lambda} = 0$ and 
$H_0 = 50$\,km\,s$^{-1}$\,Mpc$^{-1}$ 
cosmology by \citet[model C therein]{wil01}. We then 
convert this RLF to the modern cosmology adopted here, and convert
the total radio luminosity to the IC jet luminosity $L_{\gamma}$
according to equation~2. Hence, we obtain the $\gamma$-ray
luminosity function for the jets in FR~I radio galaxies, which can be
written in the form
\begin{equation}
\rho(L_{\gamma}, \, z) = \kappa (z) \times \rho_{\rm W}\left(L_{\gamma}, \, z\right) \quad ,
\end{equation}
\noindent
where $\log \rho_0 = -7.523$ (where $\rho_0$ is in Mpc$^{-3}$), $\alpha
= 0.586$, $\log L_{\rm cr} = 43.062$ (where $L_{\rm cr}$ is now the
critical intrinsic $\gamma$-ray luminosity of the jet in erg
s$^{-1}$), $z_{\rm cr} = 0.71$, and $k = 3.48$. The function $\kappa
(z)$ expresses the transformation of the comoving volume elements between
different cosmological models,
\begin{equation}
\kappa (z) \equiv {dV_{\rm W} / d\Omega \, dz \over dV / d\Omega \, dz} \quad .
\end{equation}
\noindent
We note that
\begin{equation}
{dV_{\rm W} \over d\Omega \, dz} = {c^3 \, z^2 \, (2+z)^2 \over 4 \, H_0^3 \, (1+z)^3} \quad {\rm with} \quad H_0 = 50 \, {\rm km \, s^{-1} \, Mpc^{-1}} \quad , 
\end{equation}
\noindent
and
\begin{equation}
\begin{array}{lrr}
{dV \over d\Omega \, dz} = \left({c \over H_0} \right)^3 \, E^{-1}(z) \, \left[ \int_0^z {dz' \over E(z')} \right]^2 \quad {\rm with} \quad E(z) = \sqrt{\Omega_{\rm M} \, (1+z)^3 + \Omega_{\Lambda}}, & & \\
H_0 = 71 \, {\rm km \, s^{-1} \, Mpc^{-1}} \, , \quad \Omega_{\rm M} = 0.27 \, , \quad {\rm and} \quad \Omega_{\Lambda} = 0.73 \quad .&&
\end{array}
\end{equation}
\noindent
The final form of $\rho(L_{\gamma}, \, z)$ --- shown in figure~2 (top
panel) for redshifts $z = 0.001$, $1$ and $2$ --- gives then
the number of low-power radio sources per unit comoving volume (in
Mpc$^3$) per unit (base $10$) logarithm of the intrinsic IC jet
luminosity $L_{\gamma}$ (in erg\,s$^{-1}$). We note that a
similar approach to the luminosity function was adopted by \citet{cel04},
who discussed the \emph{extended} X-ray emission of FR~I radio
galaxies (due to IC scattering of the cosmic microwave background
radiation by radio-emitting electrons) in the context of deep X-ray
surveys.

Figure~2 (bottom panel) shows also the cumulative number density 
of the low-power radio sources considered here,
\begin{equation}
\left({dN \over dV}\right)_{\rm < z} \equiv \int_0^z dz' \int_{L_{\gamma}^{\rm low}}^{L_{\gamma}^{\rm high}} {\rho(L_{\gamma}, \, z') \over L_{\gamma} \, \ln 10} \, d L_{\gamma} \quad ,
\end{equation}
\noindent
as a function of redshift, for the set of $L_{\gamma}^{\rm low} =
10^{38}$\,erg\,s$^{-1}$, $10^{39}$\,erg\,s$^{-1}$, $10^{40}$\,erg\,s$^{-1}$,
and $L_{\gamma}^{\rm high} = 10^{44}$\,erg\,s$^{-1}$. As illustrated, a
change in the lowest luminosity of the GLF by an order of magnitude
can change the number density of objects by a factor of about $3$.
Below we take $L_{\gamma}^{\rm low} = 10^{38}$\,erg\,s$^{-1}$,
corresponding to the lowest $5$\,GHz luminosity of the whole source $L_{\rm tot} = 10^{38}$\,erg\,s$^{-1}$, and note that
the lack of such low-luminosity objects in our small observed `sample'
(Table~1) is most probably due to selection effects only. However,
because of the very low $\gamma$-ray emission of such extremely
low-power radio galaxies, the particular choice of $L_{\gamma}^{\rm low}$
will not substantially affect the following estimates. On the other
hand, the number of high-luminosity sources in the population considered 
here, i.e., sources with a total radio power exceeding
Fanaroff-Riley critical luminosity (roughly $L_{\rm tot} = 10^{42}$
erg s$^{-1}$), is negligible when compared to the number of
lower-luminosity (`classical' FR~I) objects. In other words, the
adopted upper cut-off in the GLF, 
$L_{\gamma}^{\rm high} = 10^{44}$\,erg\,s$^{-1}$, can also be 
safely assumed in the following calculations. We
finally note in this context, that X-ray jets with properties
similar to the `classical' FR~I X-ray jets are indeed discovered in
radio galaxies of intermediate power and morphology \citep[e.g., 3C
15;][]{kat03}.

\section{Results and Discussion}

The $\gamma$-ray emission due
to inverse-Compton scattering of starlight photons 
in the kpc-scale jets of FR~I radio galaxies
is relatively weak, accounting for only a small fraction
of the direct EGRET measurement in the $0.1-10$\,GeV
photon energy range, consistent with the analysis presented by
\citet{cil04}. We note that for an EGRET sensitivity of 
$5.4 \times 10^{-8}$\,ph\,cm$^{-2}$\,s$^{-1}$, corresponding to a $5 \sigma$
detection in a one-year all-sky-survey \citep{blo96}, the `typical' FR~I
kpc-scale jet as considered here could be eventually detected only if
its luminosity distance is roughly $d_{\rm L} < 100 \, (L_{\gamma} /
10^{44} \, {\rm erg \, s^{-1}})^{1/2}$\,Mpc. It does not mean, however,
that the FR~I jets cannot be associated with some unidentified EGRET
sources, as the inverse-Comptonization of this and other photon fields
(especially within the inner portions of the outflows) in some particular
FR~I objects can result in $\gamma$-ray signals exceeding the
EGRET sensitivity \citep[see in this context][]{muk02,com03}.

Promisingly, GLAST is expected to be a much more powerful instrument
than EGRET, with the planned sensitivity for a $5 \sigma$ detection of
a point source within the photon energy range $0.1-300$\,GeV in
one-year all-sky-survey of $1.5 \times 10^{-9}$\,ph\,cm$^{-2}$\,s$^{-1}$
\citep{blo96}. Thus, one should expect GLAST to observe at least a few
FR~I radio galaxies at redshifts $z < 0.1$ due to the $\gamma$-ray
emission of their brightest kpc-scale knots, if the jet parameters
we assume are correct. In particular, the $dN/dz$ distribution of
the analyzed objects,
\begin{equation}
\left({dN \over dz}\right)_{\rm GL} = 4 \pi \, \int_{L_{\gamma}^{\rm GL}}^{L_{\gamma}^{\rm high}} \, {dV \over d\Omega \, dz} \, {\rho(L_{\gamma}, \, z) \over L_{\gamma} \, \ln 10} \, d L_{\gamma}
\end{equation}
\noindent
(where $L_{\gamma}^{\rm GL}$ corresponds to the minimum IC jet
luminosity which would allow for a GLAST detection as discussed
above), gives three observable FR~I jets for $L_{\gamma}^{\rm high} =
10^{44}$\,erg\,s$^{-1}$. Therefore, a GLAST survey covering low-power radio sources will be able to verify our simple model.

Meanwhile, the contribution of FR~I jets to the EGRET $\gamma$-ray
background can be evaluated for unresolved sources as
\begin{equation}
[\varepsilon I_{\gamma}(\varepsilon)] = {4 \pi \over \Omega_{\rm EG}} \, \int_{z_{\rm min}}^{\rm z_{\rm max}} {dV \over d\Omega \, dz} \, dz \, \int_{L_{\gamma}^{\rm low}}^{L_{\gamma}^{\rm high}} {\rho(L_{\gamma}, \, z) \over L_{\gamma} \, \ln 10} \, \, [\varepsilon S_{\gamma}(\varepsilon)] \, \, d L_{\gamma} \,
\end{equation}
\noindent
\citep[e.g.,][]{chi95}, where $\Omega_{\rm EG} = 10.4$ is the solid
angle covered by the survey, $[\varepsilon S_{\gamma}(\varepsilon)]$
is the observed absorbed IC flux in erg\,cm$^{-2}$\,s$^{-1}$ (equation~3), 
the comoving volume element $dV / d\Omega \, dz$ is given by the
equation~8, and the luminosity function $\rho(L_{\gamma}, \, z)$ is
given in the equation~5. The results are presented in figure~3: the
jets in FR~I radio galaxies (their brightest kpc-scale knots in
particular) can contribute about one percent to the observed
extragalactic $\gamma$-ray background in the $0.1-10$\,GeV photon
energy range. Indeed, a much larger contribution by these objects is not
expected, since accordingly to the analysis presented in, e.g.,
\citet{kne05}, blazar sources can account for the bulk of the 
extragalactic $\gamma$-ray background. It indicates that the magnetic
fields in the analyzed jet regions is not likely to be much smaller
\emph{on average} than the equipartition value, as otherwise
the contribution of the considered objects would be uncomfortably high.
Note that the IC jet luminosity scales roughly as $L_{\gamma} \propto
B^{-2}$, and hence a magnetic field as low as $10$\,$\mu$G would cause
overproduction of the diffuse $\gamma$-ray background by the FR~I jets
on their own. Therefore, weak magnetic fields in
the brightest kpc-scale knots in FR~I jets can be excluded \citep[see in this
context][for the case of M\,87 jet]{sta05}.

There are a few significant features of our analysis. First, the
analysis presented here does not depend on any particular model of particle
acceleration in the jet, as the energy distribution of the radiating
electrons is inferred directly from their synchrotron emission
\citep[see][]{sta03,sta05}. Second, we do not include here 
the inevitable
synchrotron self-Compton process, which would increase the expected IC
flux of the jets, especially in the `problematic' (because of the
absorption/re-procession effects) very high energy $\gamma$-ray band.
Note also that, accordingly to our minimum-assumption approach, we
consider the same equipartition value for the kpc-scale jet magnetic
field in \emph{all} FR~I jets, which differ in total radio power (and
hence most probably in jet kinetic luminosity) by several orders of
magnitude. Indeed, for low-power jets one should expect
an equipartition magnetic field of less than the 
$300$\,$\mu$G obtained here for
relatively powerful jets from our sample. In other words, we
underestimate the inverse-Compton radiation of such extremely weak but
numerous sources. Thus, the evaluated $\gamma$-ray emission can be
considered as a \emph{lower} limit only, and so the estimated one
percent contribution to the extragalactic $\gamma$-ray background can
be considered as \emph{guaranteed}, as long as the minimum-power
hypothesis for the kpc-scale jets in FR~I radio galaxies is correct.

Let us also mention that the presence of bright knots at $\approx
1$\,kpc from the active centers of FR~I radio galaxies is indeed
universal \citep{par87,lai99}. These knots are usually called `flaring
points', since their positions always mark the transition in jet
collimation and emissivity. \citet{lai02}, who discussed the dynamics of
the kpc-scale jet in the FR~I radio galaxy 3C~31 in detail, concluded
that the flaring points are most likely caused by the stationary
reconfinement shocks formed due to rapid changes in pressure of the
hot ambient gas \citep[see also in this context,
e.g.,][]{san83,fal85,kom94}. Others attribute them to the
Kelvin-Helmholtz instabilities, whose non-linear development can also
result in the formation of weak oblique shocks
\citep[e.g.,][]{hae79,bic96,lob03}. Whatever the case is, the analysis
presented here suggests that a weak magnetic field at the position
of the flaring points is unlikely. Of course, this conclusion is
based on a simplified approach in modeling the IC emission of the
kpc-scale FR~I jets \emph{on average}, and the jet parameters used in
our analysis are determined from a relatively small sample of the
X-ray emitting objects. However, more sophisticated analyses can be performed only
with larger sample of FR~I jets detected at optical and X-ray
frequencies.

With the approximate lower limit of $10$\,$\mu$G, the dynamical importance
of the magnetic field in kpc-scale FR~I jets cannot yet be determined,
with additional uncertainty arising from the poorly known 
total kinetic powers of these
outflows. In addition, any possible changes of the magnetic energy
flux along the outflows are not constrained in our
analysis. Let us only note in this context that modeling of the
spectral energy distribution of BL Lac objects --- believed to be
beamed counterparts of FR~I radio galaxies \citep{urr95} --- usually
leads to sub-equipartition magnetic field characterizing
the nuclear portion of the low-power jets \citep[e.g.,][]{kin02},
although the model uncertainties and variety in BL Lacs' spectral
properties are large. Hence, one can in principle suspect that
amplification of the magnetic energy flux along the low-power jets
between the sub-pc and kpc scales is required by the data, in agreement
with the more detailed analysis of M~87 \citep{sta05}. We note that
amplification of a mean magnetic energy flux is indeed expected in any
conductive plasma containing non-vanishing helicity of turbulent
motions or a strong velocity shear, although the general applicability of
the simple kinematic dynamo theory --- usually considered in this
context --- to realistic astrophysical situations (like the
relativistic outflows discussed here) is not clear \citep[see, e.g., the
recent review by][]{vis03}.

\acknowledgments

\L .S. was supported by the grants 1-P03D-011-26 and PBZ-KBN-054/P03/2001. T.M.K. was supported by the grant from the Australian Research Council. Authors acknowledge also very useful comments from P. Edwards, F. Aharonian, D. E. Harris, M. Ostrowski, and A. Siemiginowska.

\clearpage
\begin{deluxetable}{llrlrllrllll}
\tabletypesize{\scriptsize}
\tablecaption{Parameters for the brightest knots of the X-ray detected FR~I jets.}
\tablewidth{0pt}
\tablehead{
\colhead{name} 
& \colhead{$z$}
& \colhead{$d_L$}
& \colhead{$\alpha_{\rm R}$}
& \colhead{$S_{\rm R}$}
& \colhead{$\alpha_{\rm X}$}
& \colhead{$S_{\rm X}$}
& \colhead{$S_{\rm O}$}
& \colhead{$\log$ $\nu_{\rm br}$}
& \colhead{$B_{{\rm eq}}(1)$}
& \colhead{$\log$ $L_{\rm tot}$}
& \colhead{$\eta$}\\  
& & [Mpc] & & [mJy] &  & [nJy] & [$\mu$Jy] & [Hz] & [$\mu$G] & [erg s$^{-1}$] & }
\startdata
3C 15   & 0.073   & 326 &  0.85 &   55 & 1.2 &  1.2 & 1.6 
& 14.16 & 877 & 42.0  & 0.035 \\ 
NGC 315 & 0.0165  &  71 &  0.90 &   68 & 1.2 &  4.1 & --- 
& 16.37 & 392 & 40.6 & 0.047 \\
3C 31   & 0.0169  &  72 &  0.55 &   37 & 1.2 &  7.3 & 2.0 
& 13.57 & 334 & 40.8 & 0.02 \\
B 0206  & 0.0369  & 160 &  0.50 &   26 & 1.2 &  5.2 & --- 
& 13.30 & 474 & 41.1 & 0.033 \\
3C 66B  & 0.0215  &  92 &  0.60 &   34 & 1.2 &  6.2 & 15.8
& 13.84 & 374 & 41.3 & 0.01 \\
3C 129  & 0.0208  &  89 &  0.75 &  3.8 & 1.2 &  1.9 & --- 
& 16.19 & 196 & 41.1 & 0.001 \\
B2 0755 & 0.0428  & 187 &  0.75 &   54 & 1.2 &  9.7 & ---
& 15.20 & 637 & 41.1 & 0.091 \\
M 84    & 0.00354 &  15 &  0.65 &   13 & 1.2 & 1.16 & $<$30 
& 13.65 & 101 & 39.9 & 0.002 \\
M 87    & 0.00427 &  18 &  0.70 & 2600 & 1.6 &  142 & 1000
& 15.29 & 510 & 41.3 & 0.024 \\
Cen A   & 0.00081 & 3.4 &  0.75 &  520 & 1.5 &  110 & ---
& 16.17 & 125 & 40.8 & 0.0005 \\
3C 296  & 0.0237  & 102 &  0.60 & 23.9 & 1.4 &  1.2 & $<$10  
& 14.02 & 358 & 41.0 & 0.015 \\
\enddata
\tablecomments{ 
$z$: redshift of the source,
$d_{\rm L}$: luminosity distance to the source adopting 
$H_0$ = 71 km s$^{-1}$ Mpc$^{-1}$, $\Omega_{\rm m}$ = 0.27 and 
$\Omega_{\Lambda}$ = 0.73.     
$\alpha_{\rm R}$: radio spectral index at 5\,GHz, 
$S_{\rm R}$: radio flux density at 5\,GHz in mJy,  
$\alpha_{\rm X}$: X-ray spectral index at 1 keV, 
$S_{\rm X}$:  X-ray flux density at 1\,keV in nJy,
$S_{\rm O}$:  optical flux density at 5$\times$10$^{14}$\,Hz in $\mu$Jy,
$\nu_{\rm br}$: break frequency of the synchrotron emission in Hz,
$B_{\rm eq}(1)$: the equipartition magnetic field computed for no relativistic beaming in $\mu$G,
$L_{\rm tot}$: total luminosity of the source at 5 GHz in erg s$^{-1}$, and  
$\eta$: ratio between 5 GHz luminosity of the knot and total 5 GHz luminosity of the source. {\bf References ---} 3C 15: \citet{kat03}; NGC 315: \citet{wor03}; 3C 31: \citet{hard02}; B 0206: \citet{wor01}; 3C 66B: \citet{hard01}; 3C 129: \citet{har02a}; B2 0755: \citet{wor01}; M 84: \citet{har02b}; M 87: \citet{mar02,wil02}; Cen A: \citet{kra02,hard03}; 3C 296: \citet{hard05}. Total radio luminosities are taken from \citet{liu02} and converted to the adopted cosmology and an observing frequency of 5\,GHz.}
\end{deluxetable}
\clearpage
\begin{figure}
\includegraphics[scale=1.50]{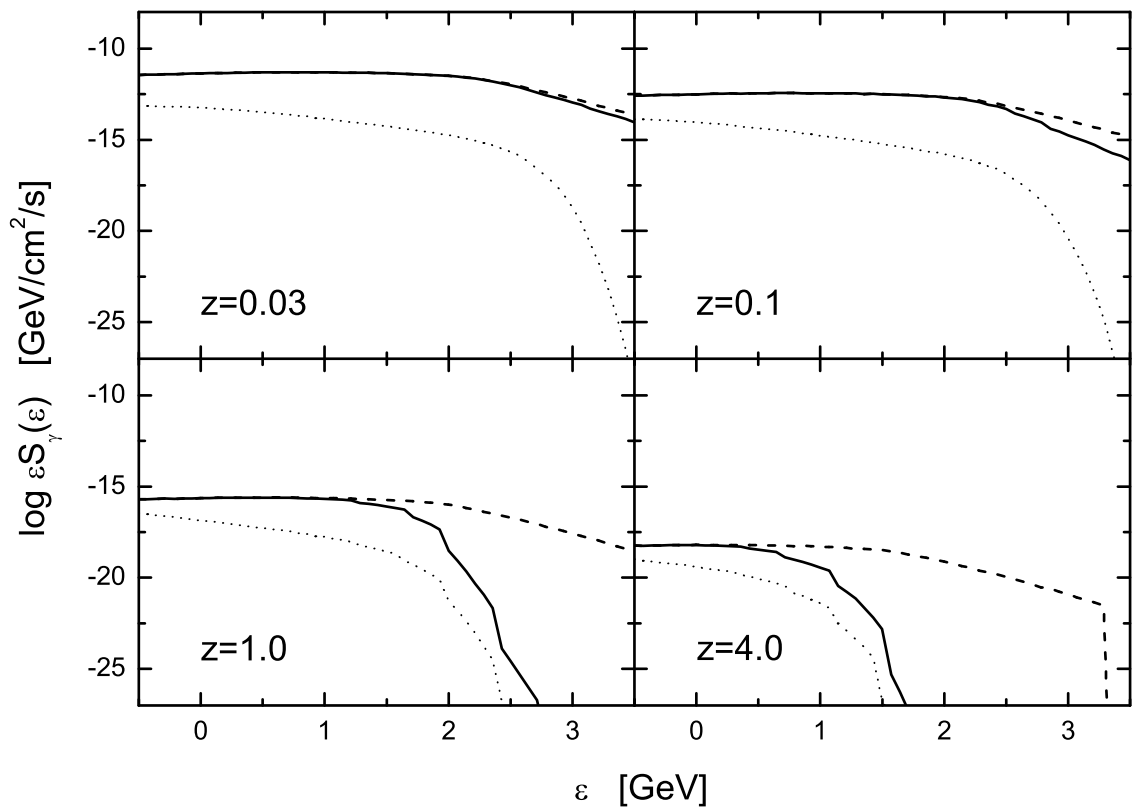}
\caption{
$\gamma$-ray spectra of kpc-scale FR~I jets located at 
redshifts $z=0.03$, $0.1$, $1.0$ and $4.0$ (as indicated at each
panel) for a total IC jet luminosity $L_{\gamma} = 10^{41}$\,erg\,s$^{-1}$.
Dashed lines indicate emission intrinsic to the source, thick solid
lines correspond to the emission which would be measured by the observer located at $z = 0$ (with absorption/re-emission
effects included), while dotted lines illustrate emission from the
source's halo.}
\end{figure}

\begin{figure}
\includegraphics[scale=1.50]{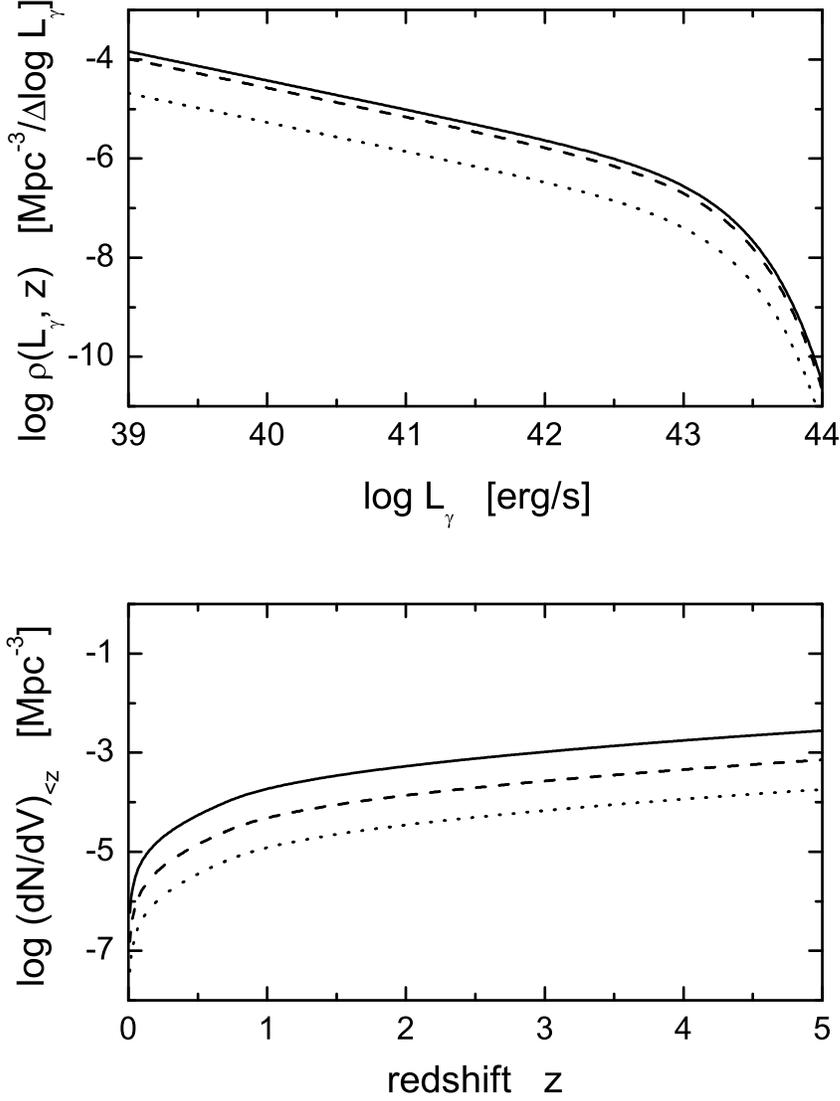}

\caption{
{\bf {\it top:}} GLF of kpc-scale FR~I jets for a modern cosmology
($\Omega_{\rm M} = 0.27$, $\Omega_{\Lambda} = 0.73$, 
$H_0 = 71$\,km\,s$^{-1}$\,Mpc$^{-1}$) 
and redshifts $z=0.001$, $1$, and $2$
(dotted, dashed and solid lines, respectively) as a function of total
IC jet luminosity $L_{\gamma}$. {\bf {\it bottom:}} Cumulative number
density of low-power radio sources as a function of redshift, for
$L_{\gamma}^{\rm low} = 10^{38}$\,erg\,s$^{-1}$, $10^{39}$\,erg\,s$^{-1}$
and $10^{40}$\,erg\,s$^{-1}$ (solid, dashed and dotted lines,
respectively), and $L_{\gamma}^{\rm high} = 10^{44}$\,erg\,s$^{-1}$.}

\end{figure}

\begin{figure}
\includegraphics[scale=1.50]{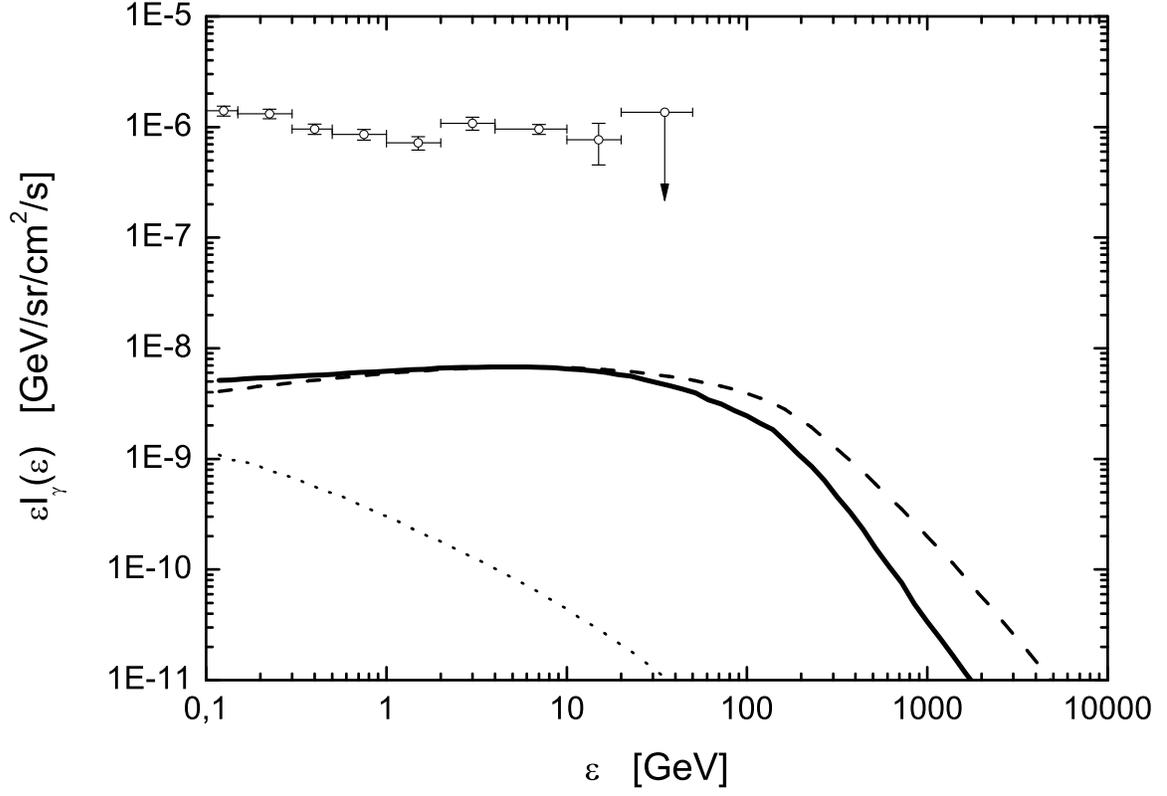}
\caption{
Contribution of FR~I kpc-scale jets to the extragalactic $\gamma$-ray
background for $z_{\rm min} = 0$, $z_{\rm max} = 5$, $L_{\gamma}^{\rm low} = 10^{38}$\,erg\,s$^{-1}$, and 
$L_{\gamma}^{\rm high} = 10^{44}$\,erg\,s$^{-1}$ 
(lines denoted as in figure~1). Open circles correspond
to the extragalactic EGRET $\gamma$-ray background as determined by
\citet{str04}.}
\end{figure}

\end{document}